\documentclass[prl,amsmath,amssymb, aps,onecolumn,10pt,superscriptaddress]{revtex4-2}
\usepackage{graphicx,xcolor,array,float,amsmath,amsfonts}
\usepackage[utf8]{inputenc}
\usepackage{graphicx}  
\usepackage{float}     
\usepackage{amsmath}   

\usepackage{xr}
\makeatletter
\newcommand*{\addFileDependency}[1]{
  \typeout{(#1)}
  \@addtofilelist{#1}
  \IfFileExists{#1}{}{\typeout{No file #1.}}
}
\makeatother
\newcommand*{\myexternaldocument}[1]{
    \externaldocument{#1}
    \addFileDependency{#1.tex}
    \addFileDependency{#1.aux}
}
\myexternaldocument{SI}

\begin{document} 

\title{Seeing inside a Plasmonic Nanogap: Few-molecule Orientation and Preferential Adsorption}
\author{Francesco Ciccarello}
\author{Nobuaki Oyamada} 
\author{Thwahira Shirin Alampara} 
\author{Konstantin Malchow}
\affiliation{Institute of Physics, Ecole Polytechnique F\'ed\'erale de Lausanne (EPFL), CH-1015 Lausanne, Switzerland}
\author{Christophe Galland}
\email{chris.galland@epfl.ch}
\affiliation{Institute of Physics, Ecole Polytechnique F\'ed\'erale de Lausanne (EPFL), CH-1015 Lausanne, Switzerland}
\affiliation{Center for Quantum Science and Engineering, Ecole Polytechnique F\'ed\'erale de Lausanne (EPFL), CH-1015 Lausanne, Switzerland}

\begin{abstract}
\textbf{Molecule-surface interactions are central to many research and technological areas, spanning from heterogeneous catalysis and polymer science to electrochemistry. Of particular relevance are metallic nanogaps used in molecular electronics and near-field spectroscopy. Due to the buried nature of these double interfaces, few methods exist to monitor side-specific interactions and relative molecular orientation inside the gap. 
In this work, we introduce plasmon-enhanced nonlinear vibrational spectroscopy as an efficient tool to investigate surface molecular adsorption within metallic nanojunctions. 
By exploiting simultaneous vibrational sum- and difference-frequency generation in dual-resonant nanocavities, we resolve molecular orientation and preferential binding to one of the two gold surfaces, with few-molecule sensitivity.
We also discover that the non-resonant (electronic) second-order nonlinear response is not an intrinsic property of the metal surface, but is instead governed by the molecule-surface interaction.
Our findings provide a powerful analytical tool, easily implementable as an add-on to Raman spectroscopy, thanks to commercially available mid-infrared quantum cascade lasers. 
}
\end{abstract}
\maketitle

\section{Introduction}\label{sec1}
Understanding and controlling surface chemistry near metals is crucial to preserve and tune the properties of numerous systems, such as organic solar panels and LEDs, hydrophobic coatings, reactive electrodes, and a variety of sensors \cite{liuSuperHydrophobicIcephobicCoatings2016,schoenbaumControllingSurfaceEnvironment2014,yiSelfassembledMonolayersSAMs2024,chakiSelfassembledMonolayersTunable2002}. Most of these processes are governed by physical mechanisms at the nanometer scale, making their dynamics difficult to probe: conventional surface characterization techniques such as X-ray photoelectron spectroscopy (XPS) \cite{blomfieldSpatiallyResolvedXray2005a}, water-contact angle measurements \cite{thorneContactAngleMapping2025} and Auger electron spectroscopy (AES) provide useful information on large ensembles, but they often lack microscopic spatial resolution and few-molecule sensitivity. Conversely, scanning tunneling microscopy (STM), atomic force microscopy (AFM), and tip-enhanced Raman spectroscopy (TERS) can achieve single-molecule sensitivity and nanoscale lateral resolution \cite{jiangObservationMultipleVibrational2012,klingspornIntramolecularInsightAdsorbate2014}, but they are unable to probe the inside of metallic nanogaps, and usually do not provide direct information on molecular orientation.

\begin{figure}[H] 
        \centering
        \includegraphics[width=1\linewidth]{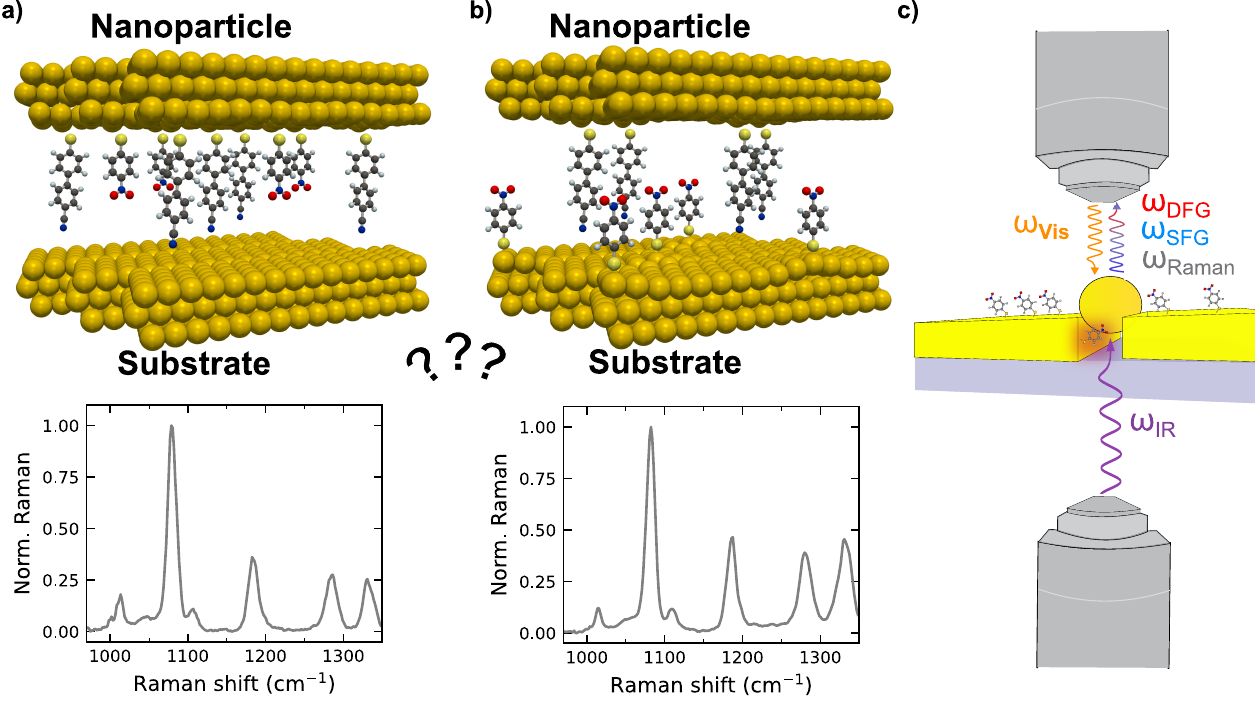}
        \caption{\textbf{(a)} Sketch of a gold-molecule-gold nanogap with 4-cyanobiphenylthiol (CN-BPT) and 4-nitrophenylthiol (4NTP) forming a mixed SAM on the nanoparticle. The bottom plot shows the measured SERS spectrum. \textbf{(b)} Sketch of a hybrid nanogap where the two molecules are on opposite sides. The SERS spectrum, in the bottom plot, shows no distinctive features compared to (a). \textbf{(c)} Concept of the experiment. 
        The probed molecules are trapped between the nanoparticle and the slit walls, where both the mid-infrared and visible fields are enhanced. }
        \label{fig:concept}
    \end{figure}

Surface-enhanced Raman spectroscopy (SERS) has recently emerged as an easily deployable tool to probe the dynamics of small molecular ensembles \cite{liuSpatiotemporalRamanProbing2026}, and its limit has been pushed down to the single molecule level \cite{park2010charge} by exploiting randomly occurring hot spots  \cite{kneippSingleMoleculeDetection1997,etchegoinPerspectiveSingleMolecule2008} and picocavities in nanoresonators \cite{baumbergPicocavitiesPrimer2022,griffithsLocatingSingleAtomOptical2021}, where the effective optical mode volume is believed to reach a few cubic nanometers only. However, using SERS alone, 
it is challenging to resolve molecular orientation and surface-selective binding in symmetric nanogaps \cite{chulhaiDeterminingMolecularOrientation2013,akbaliDeterminingMolecularOrientation2021}, as shown in Figs.~\ref{fig:concept}a-b.
   
On the other hand, vibrational sum-frequency generation (vSFG) and external reflectance Fourier transform infrared (ER-FTIR) spectroscopy on plasmonic substrates can provide information about the orientation of localized vibrations, but are typically limited to large molecular ensembles and single interfaces \cite {valleyStericHindrancePhotoswitching2013,kangMixedSelfAssembledMonolayers1999,yeganehInterfacialAtomicStructure1995}.
The relative molecule-surface orientation was determined using ultrafast (with pico- or femto-second laser pulses) vSFG on homogeneous monolayers by addressing localized vibrations, such as nitro or nitrile group stretching modes. Retrieving the group orientation requires several measurements with different polarization combinations \cite{darwishCharacterizingPhotoinducedSwitching2012}, advanced fitting algorithms that rely on quantities that are a priori unknown \cite{sovagoDeterminingAbsoluteMolecular2009}, or phase-sensitive measurements with interferometric schemes and a local oscillator \cite{balosPhaseSensitiveVibrationalSum2022}.

Recent efforts in vSFG spectroscopy have been dedicated to probing smaller molecular ensembles; first, in densely packed gold nanoparticle monolayers, where numerous hotspots are sampled at the same time \cite{humbertSumFrequencyGenerationSpectroscopy2019,zhengSinglemoleculelevelDetectionInterfacial2025}, then, in individual nanocavities engineered to support colocalized mid-infrared and visible resonances \cite{xomalisDetectingMidinfraredLight2021,chenContinuouswaveFrequencyUpconversion2021,xieContinuousWaveHighResolutionUltrabroadband2026} or single NPoM trough tip-enhanced vSFG (TE-vSFG) \cite{roelliInoperandoControlSumfrequency2025}. 
Latest advances in ultrafast TE-vSFG have pushed the resolution down to approximately a hundred molecules on a flat gold surface, and pulse shaping has helped retrieve molecular orientation \cite{sakuraiTipenhancedSumFrequency2026}, but its applicability on buried structures and nanojunctions has not been tested yet, and molecular signal degradation remains a serious issue under pulsed excitation \cite{crampton2016ultrafast}.
 
Here, we present a method that combines sub-wavelength probe volume, easily deployable continuous-wave excitation, and few-molecule sensitivity, which allows for unambiguous determination of relative molecular orientation inside a nanogap.
The method builds on a dual-resonant nanocavity design allowing straightforward and simultaneous access to vSFG and vDFG over a broad wavenumber range (900-1650 cm$^{-1}$), free of geometric phase-matching  (Fig.~\ref{fig:concept}c)  \cite{xieContinuousWaveHighResolutionUltrabroadband2026}. 
We process these signals to infer the relative orientation of molecular adsorbates in the gap, leveraging the interference between the resonant and non-resonant contributions to the second-order $\chi^{(2)}$ response of the metal-molecule-metal assembly. The method is validated by preparing hybrid nanogaps with two molecular species selectively anchored on either side and retrieving their relative orientation from vSFG/vDFG measurements. 

With this approach, we discover that the non-resonant $\chi^{(2)}$ response, commonly seen as an intrinsic property of the metal surface where inversion symmetry is broken, is in fact dominated by the molecule-surface interaction. The binding of a molecule to the metal surface strongly alters its second-order nonlinear response, possibly due to the formation of a charge-transfer state and to the atomic restructuring of the surface \cite{yeganehInterfacialAtomicStructure1995,bussonNonUniquenessParametersExtracted2009}. 
Finally, we track the fate of molecules added after the assembly of the nanogap, which diffuse inside it \cite{liuSpatiotemporalRamanProbing2026,goerlitzerSurfaceSelectiveMolecularBinding2026} and determine the preferential binding sites of these molecules without prior knowledge, in a single experimental realization. 



\section*{Theoretical background}
We first summarize the main theoretical principles of vSFG and vDFG \cite{linMolecularTheorySecondorder1996}.
Their intensities are proportional to the square modulus of the effective (contracted tensor) second-order susceptibility $|\chi^{(2)}_{\rm eff}(\pm\omega)|^2$, which describes the nonlinear optical response of the material as a function of the mid-infrared (MIR) excitation frequency $\omega$ (we keep the visible pump frequency fixed). In centrosymmetric media, under the dipole approximation, $\chi^{(2)}(\pm\omega)$ is zero; thus, in the bulk material, SFG and DFG are forbidden. However, due to symmetry breaking, $ \chi^{(2)}(\pm\omega) $ is nonzero at the surface, making these techniques intrinsically surface-sensitive.
At the metal-molecule interface, the nonlinear response can be divided into two contributions:
\begin{equation}
|\chi^{(2)}_{\mathrm{eff}}(\pm\omega)|^2 \propto \left|\chi^{(2)}_{\mathrm{NR}}+\chi^{(2)}_\mathrm{Res}(\pm\omega)\right|^2 \label{eq:modulus}
\end{equation} 
Here $\chi^{(2)}_{\rm NR}$  is a non-resonant (MIR frequency-independent) contribution resulting from all far-detuned electronic transitions. It is commonly associated with the metal surface, irrespective of the presence of the molecule, and also contributes to second-harmonic generation for any incoming frequency \cite{meier2023controlling}. The resonant contribution $\chi^{(2)}_{\rm Res}$ models the response of molecular vibrations that are both Raman and IR-active. Treating these vibrations as harmonic oscillators, eq.~\eqref{eq:modulus} can be made explicit for the induced polarisations at the SFG and DFG frequencies:
\begin{equation}
\chi^{(2)}_{\mathrm{eff}}(+\omega) = \sum_{j=1}^{N} \frac{A_j e^{i\phi_j}}{\omega_j - \omega - i\gamma_j} + A_{\mathrm{NR}} e^{i\phi_{\mathrm{NR}}} \label{eq:SFG}
\end{equation}
\begin{equation}
\chi^{(2)}_{\mathrm{eff}}(-\omega) = \sum_{j=1}^{N}  \ \frac{A_j e^{i\phi_j}}{\omega_j - \omega + i\gamma_j} + A_{\mathrm{NR}} e^{i\phi_{\mathrm{NR}}} \label{eq:DFG}
\end{equation}
where $\phi_{\rm NR}$ and $A_{\rm NR}$ are the phase and amplitude of $\chi^{(2)}_{\rm NR}$, while $\omega_j$, $\phi_j$, $\gamma_j$ and $A_j$ are, respectively, the frequency, phase offset, linewidth, and amplitude of the $j^{th}$ vibrational mode of the molecule. In a simplified one-dimensional model, this amplitude is proportional to the product of the IR transition dipole moment $\frac{\partial\mu}{\partial Q_j}$ and the Raman polarizability$\frac{\partial \alpha}{\partial Q_j}$. 
In the following, frequencies and linewidths are expressed in wavenumber units [cm$^{-1}$], where $\omega~[\text{rad/s}] = 200\pi c \ \nu~[\text{cm}^{-1}]$.  
SFG and DFG are expected to share almost the same $\phi_{\rm NR}$ phase in our experimental conditions, because under 780~nm excitation, both outgoing frequencies are far from any interband transitions in gold \cite{dalsteinNonlinearOpticalResponse2018} (we discuss later the possible role of charge-transfer resonances). 
On the contrary, the imaginary part of the resonant nonlinear susceptibility flips signs between the SFG and DFG frequencies: $\Im{[\chi^{(2)}_{\mathrm{Res}}(+\omega)]}=-\Im{[\chi^{(2)}_{\mathrm{Res}}(-\omega)]}$, while the real part is symmetric: $\Re{[\chi^{(2)}_{\mathrm{Res}}(+\omega)]}=\Re{[\chi^{(2)}_{\mathrm{Res}}(-\omega)]}$.

\begin{figure}[t] 
        \centering
        \includegraphics[width=1\linewidth]{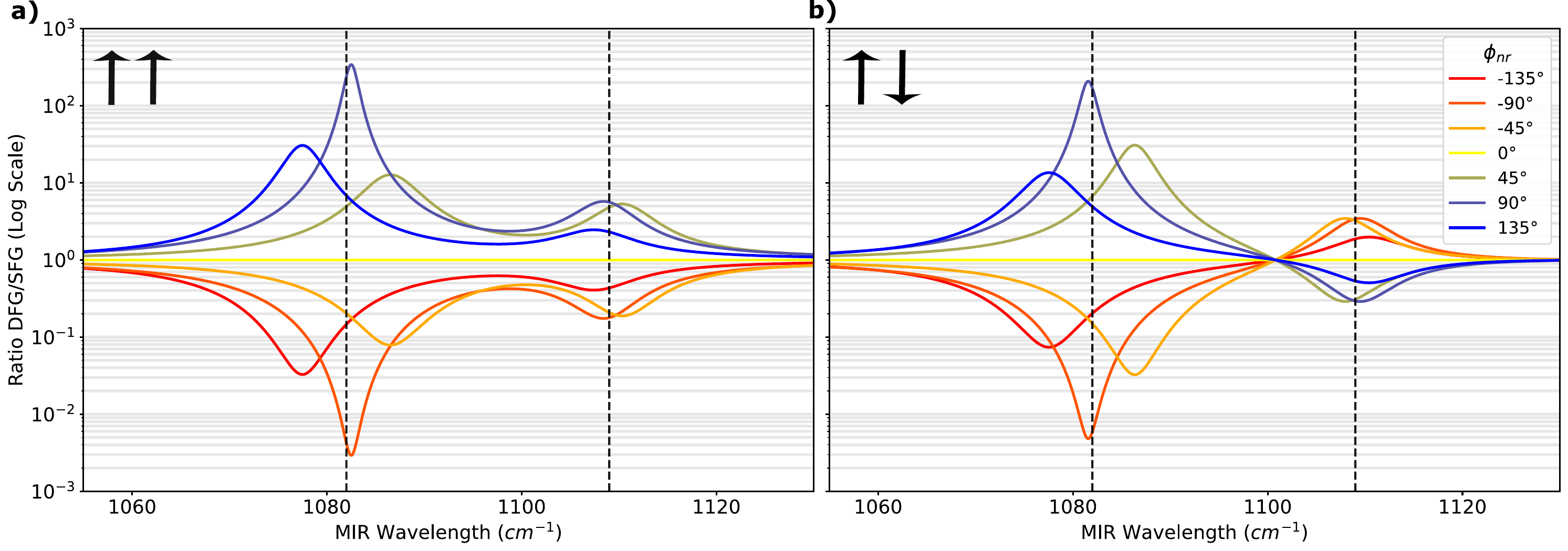}
        \caption{\textbf{(a)} $\phi_{\rm NR}$ dependence of simulated DFG/SFG ratio $R(\omega)$ for two vibrations which share the same phase offset $\phi_1=\phi_2=0$. The modes at 1080 and 1109~cm$^{-1}$ match the CN-BPT and 4NTP vibration of Fig.~4. \textbf{(b)} Same, but the 1109~cm$^{-1}$ mode has now $\phi_2=\pi$, as when the two molecules are anti-parallel. Vertical dashed lines indicate the two vibrational frequency. 
        The following parameters, extracted from Fig.~4, have been used in the model: $\gamma_1 = 8.0~\mathrm{cm}^{-1}$, $\gamma_2 = 6.0~\mathrm{cm}^{-1}$, $A_1 = 7.1$, $A_2 = 1.7$ and $A_{NR} = 1$.}
        \label{fig:theory}
    \end{figure}

We now consider the ratio $R(\omega)=\left|\chi^{(2)}_{\rm eff}(-\omega)\right|^2 / \left|\chi^{(2)}_{\rm eff}(+\omega)\right|^2$, plotted in Fig.~\ref{fig:theory} for a toy model consisting of just two molecular vibrations with resonant frequencies marked by dashed lines. $R(\omega)$ is directly accessible with our setup and is robust to power fluctuations and sample drift \cite{xieContinuousWaveHighResolutionUltrabroadband2026}. 
In the left panel, the two vibrations have the same relative phase offset $\phi_j=0$, corresponding to parallel molecules. In the right panel, one vibration has a $\pi$ phase offset, corresponding to anti-parallel molecules. 
Because of the symmetry of $\chi^{(2)}_\mathrm{Res}$ between SFG and DFG about the real axis in the complex plane, we see that if $\chi^{(2)}_{\mathrm{NR}}$ is purely real ($\phi_{\rm NR}\simeq 0$), then the ratio $R(\omega)$ is constant and insensitive to molecular orientation, as confirmed by the yellow curves in Figs.~\ref{fig:theory}a,b. 
In each panel, the other curves correspond to $\phi_{\rm NR}$ being varied as indicated in the legend: 
the larger the imaginary part of $\chi_{\rm NR}$, the stronger the variation in $R(\omega)$ around each vibrational resonance. 
These variations follow the same sign for the two modes in the case of parallel molecules, but are opposite for anti-parallel molecules. 
We also see that the extrema of $R(\omega)$ do not coincide, in general, with the Raman shifts of the respective vibrations \cite{bussonNonUniquenessParametersExtracted2009,pattersonNonresonantSumfrequencyGeneration2024}.
This is a consequence of the combined interference with $\chi_{\rm NR}$ and with the other vibrations. 
We now apply these lessons to the interpretation of experimental data. 

\section*{Results}
\paragraph{Measurement of $R(\omega)$ at the nanoscale}
The dual-resonant nanoparticle-on-slit (NPoS) cavity concentrates both MIR and visible light into a pair of nanogaps between a metallic nanoparticle (gold here) and the sidewalls of the metal slit (also gold), increasing the optical nonlinear response from the few molecules in this gap. We patterned dense arrays of slits using electron beam lithography and ion beam etching, which work as MIR metasurfaces with resonance frequencies defined by the slit length and spacing \cite{xieContinuousWaveHighResolutionUltrabroadband2026}. Here we chose these parameters to have the MIR resonance centered at 1100 cm$^{-1}$.
A 780~nm, weak continuous-wave visible laser beam (a few $\mu$W on the sample) is focused on the gold surface through a high numerical aperture (NA=0.9) objective. A tunable MIR QCL laser is focused by a Cassegrain objective (NA=0.78) through the silicon substrate on the same point. The MIR power on the slit is between 1 and 10 mW, with a larger spot size of 10 to 15 $\mu$m in diameter. 
When the two beams are focused on an NPoS, the vSFG, vDFG, and SERS signals from the molecules are measured simultaneously on the spectrometer after blocking the reflected laser light with a notch filter. Stepwise tuning of the MIR wavelength allows vSFG and vDFG spectra to be acquired with 1~cm$^{-1}$ resolution (independent of spectrometer resolution and calibration) over several hundreds of cm$^{-1}$.

\begin{figure}[t]
        \centering
        \includegraphics[width=1\linewidth]{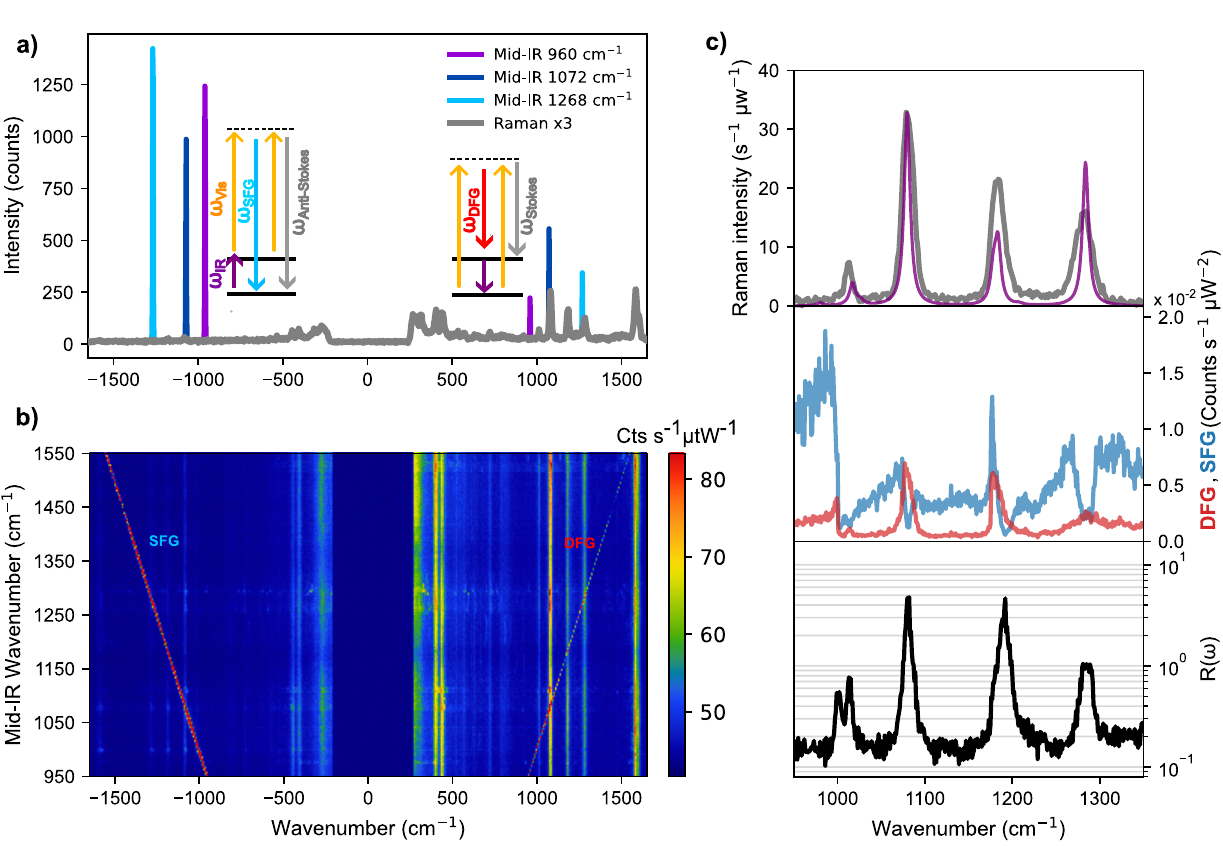}
        \caption{ \textbf{(a)} Emission spectra
from a single CN-BPT SAM functionalized NPoS obtained at three different MIR frequencies (light blue, blue, and purple lines), and without MIR
pump (SERS only, grey curve). Energy diagrams are shown in the inset for v-SFG, v-DFG, and the Raman process. The VIS power is 3 $\mu$W on the sample. \textbf{(b)} Color map of a full MIR scan of the same NPoS from 950 to 1660 cm$^{-1}$. Oblique lines correspond to SFG and DFG signals. Counts are normalized by VIS power. \textbf{(c)} The top plot shows the Stokes SERS (grey curve) together with the DFT-calculated CN-BPT spectrum (purple line). In the middle graph, DFG (red) and SFG (blue) spectra are plotted, normalized by the product of VIS and IR power. The bottom graph shows their logarithmic ratio (black line). The exposure time is 2 s for each MIR wavenumber. }
        \label{fig:spectra}
\end{figure}

In Fig.~\ref{fig:spectra}a, we show representative emission spectra from an NPoS hosting a 4-cyanobiphenylthiol (CN-BPT) self-assembled monolayer (SAM) acting as a molecular spacer, for 3 different MIR excitation wavelengths. 
On top of spontaneous Raman scattering, sharp coherent DFG and SFG emission peaks appear on the Stokes and anti-Stokes side of the spectrum, respectively (colored curves). 
When the MIR wavenumber is tuned, a two-dimensional dataset is acquired, as illustrated in Fig.~\ref{fig:spectra}b. 
For each MIR wavenumber, we extract the net DFG and SFG intensities, thereby obtaining the spectra of Fig.~\ref{fig:spectra}c.
Their ratio $R(\omega)$ is plotted on a logarithmic vertical scale as a black curve. 
Its flat baseline confirms that $R(\omega)$ is robust against laser fluctuation and sample drift, and is insensitive to the NPoS MIR resonance profile \cite{xieContinuousWaveHighResolutionUltrabroadband2026}.

A set of prominent peaks in $R(\omega)$, slightly shifted from the SERS peaks, and corresponding to the IR+Raman-active vibrational modes of CN-BPT, indicate that the non-resonant phase $\phi_{\rm NR}$ has a large imaginary part, as discussed in Fig.~\ref{fig:theory}; hence, analysis methods based on the assumption that $\Im[\phi_{\rm NR}]=0$ are not applicable \cite{pattersonNonresonantSumfrequencyGeneration2024,sunPhaseSensitiveSumFrequencyGeneration2019}. 
The fact that $\chi^{(2)}_{\rm NR}$ should be purely real for metals has been recently questioned in multiple works \cite{pattersonNonresonantSumfrequencyGeneration2024,fellowsObtainingExtendedInsight2023,liHeterodyneTransientVibrational2019,sunPhaseSensitiveSumFrequencyGeneration2019}. Concerning gold surfaces, there is no agreement among values present in the literature \cite{dreesenInfluenceMetalElectronic2002,plucheryEnhancedDetectionThiophenol2009,fellowsObtainingExtendedInsight2023,liHeterodyneTransientVibrational2019,marmolejosGoldStandardPhase2019} and proper correction must be applied to avoid artifacts \cite{poolComparativeStudyDirect2011}.
More routinely, the phase information, which is fundamental to retrieving molecular orientation, is extrapolated by fitting the nonlinear optical response. But even for a single vibrational mode, two different sets of $\phi_{\rm NR}$ and amplitude describe equivalently the vSFG response \cite{plucheryEnhancedDetectionThiophenol2009}. For a richer spectrum, like in Fig.~\ref{fig:spectra}c, the degeneracy of the fit parameters increases, and the situation is even worse when a phase offset is introduced for each mode, as explained in \cite{bussonNonUniquenessParametersExtracted2009}. 
In our experiment, synchronous recording of vSFG and vDFG helps to lower the degeneracy of possible solutions \cite{bussonNonUniquenessParametersExtracted2009}.
Moreover, due to the field localization in the nanogaps, we can consider that only the $zzz$ component of the nonlinear tensor is contributing to $\chi_{\rm eff}^{(2)}$, which in principle rules out values other than 0 or $\pi$ for the phase offset of each mode \cite{bussonNonUniquenessParametersExtracted2009}.


\paragraph{Molecular orientation in the nanogap}

\begin{figure}[t]
        \centering
        \includegraphics[width=.9\linewidth]{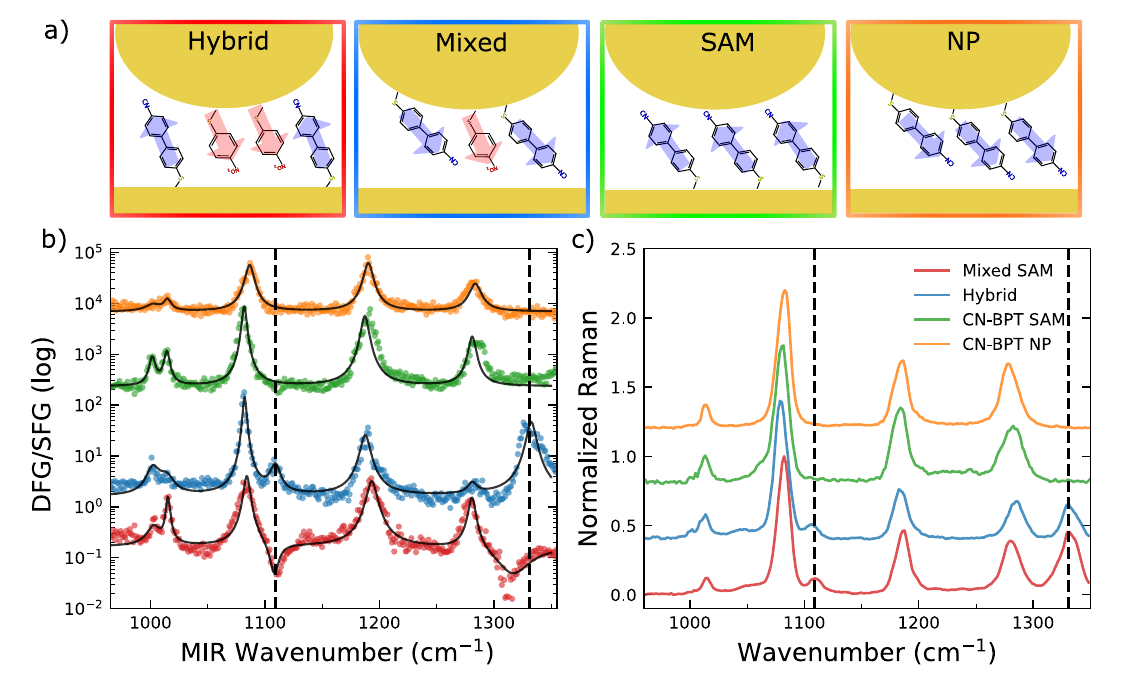}
        \caption{\textbf{(a)} Sketch of the four molecular configurations described in the main text, with the color code used in the following panels and arrows to highlight molecular orientation. \textbf{(b)} DFG/SFG intensity ratio $R(\omega)$ plotted on a logarithmic scale for each configuration of panel (a). Thin black lines are fits according to eqs.~(\eqref{eq:SFG}) and (\eqref{eq:DFG}) (fit parameters listed in Table~\ref{tab:fit_parameters_deg}). \textbf {(c)} Respective SERS spectra of the NPoS of panel (b), normalized to the peak intensity at 1082 cm$^{-1}$. Dashed vertical lines mark the vibrational frequencies characteristic of 4NTP only.}
        \label{fig:orientation}
    \end{figure}

We now analyze four different types of NPoS and demonstrate how to resolve the molecular orientation, thereby also pinpointing the spatial origin of $\chi^{(2)}_{\rm NR}$. In addition to CN-BPT, we introduce 4-nitrophenylthiol (4NTP) as a second molecular reporter, which has two distinct vibrational modes around 1109 and 1330~cm$^{-1}$ non-overlapping with CN-BPT. 
The four samples investigated in Fig.~\ref{fig:orientation} consist of: (i) hybrid constructs with a 4NTP SAM on the gold slit and a functionalized CN-BPT nanoparticle on top of it [named `hybrid']; (ii) mixed SAM of CN-BPT and 4NTP formed on the gold nanoparticle (NP) before deposition in the slit [named `mixed']; (iii) pure CN-BPT SAM on the gold slit; and (iv) pure CN-BPT SAM on the nanoparticle. 
The ratio CN-BPT:4NTP in the mixed SAM sample is 3:1, and has been evaluated through SERS intensity calibration according to \cite{bellCoherentDynamicsMolecular2025,muellerCollectiveMidInfraredVibrations2022} (see Fig.~\ref{fig:SICalibration}).

The first key observation is that the two pure CN-BPT samples show qualitatively similar spectra, although the molecular orientation in the nanogap is flipped between them. It is established that reversing the molecular orientation incurs a $\pi$ phase shift to the resonant $\chi^{(2)}$ response \cite{darwishCharacterizingPhotoinducedSwitching2012}. Consequently, if the nonresonant $\chi^{(2)}$ response were coming from a specific gold surface (e.g., that of the nanoparticle), independent of the molecular binding, the orange and green curves of Fig.~\ref{fig:orientation}b would be the mirror images of each other, with peaks replaced by dips.  
Hence, our experimental findings reveal that upon changing the molecular orientation in the gap, the nonresonant response changes sign. Molecular binding thus strongly enhances the nonresonant response of the interface.

As emphasized by Patterson \cite{pattersonNonresonantSumfrequencyGeneration2024}, the fundamental origin of $\chi^{(2)}_{\rm NR}$ remains poorly understood. 
Previously, for SAM on metallic surfaces, Dreesen et al. reported a significant dependence of $\chi^{(2)}_{\rm NR}$ on the specific metal substrate \cite{dreesenInfluenceMetalElectronic2002}, attributed to the varying interplay between inter-band and intra-band transitions. In our case, the energy of 780~nm photons is far detuned from interband transitions in gold.
But, in contrast with Ref.~\cite{dreesenInfluenceMetalElectronic2002} the molecule employed in this work are much shorter and their LUMO is much lower in energy, which could enable optical excitation of metal-to-ligand charge transfer (MLCT) states contributing to $\chi^{(2)}_{\rm NR}$. 
Additionally, in our nanogap geometry, the  $\chi^{(2)}_{\rm NR}$ contributions of the two facing gold surfaces may partially cancel out, resulting in a regime where the contribution of MLCT states to $\chi^{(2)}_{\rm NR}$ stands out. This finding motivates experiments to quantify this contribution \cite{de2024charge}.

We now focus on the mixed and hybrid NPoS: in the former, CN-BPT and 4NTP are parallel, in the latter, they are anti-parallel (with respect to their gold binding sites). 
We find that the nonresonant response of the gold surface functionalized with 4NPT is much weaker than with CN-BPT (see also Fig.~\ref{fig:SI4NTP}) \cite{xieContinuousWaveHighResolutionUltrabroadband2026,arulCoherentSumfrequencyGeneration2026}, so that the modes of CN-BPT still appear as peaks in both spectra (red and blue curves in Fig.~\ref{fig:orientation}b). 
Another striking feature is the reversal of the 4NTP resonances from peaks to dips when these molecules are anti-parallel to CN-BPT. 
It is precisely the expected behaviour from the theory discussed in Fig.~\ref{fig:theory}, and confirms the sensitivity and accuracy of our method to probe molecular orientation in nanogaps in a non-invasive way. The reproducibility of our findings is illustrated in Figs.~S2-S5 with several spectra obtained from different NPoS for each gap configuration.  
We also fabricated an opposite hybrid NPoS compared to (i), where 4NTP is on the NPs and CN-BPT on the slit: as shown in Fig.~\ref{fig:SIopposite}, its $R(\omega)$ spectrum shows the same peaks and dips as configuration (i), strengthening our conclusion that $\chi^{(2)}_{NR}$ is chiefly governed by molecule-surface interactions.

\paragraph{Application to diffusing molecules}

\begin{figure}[t]
        \centering
        \includegraphics[width=.8\linewidth]{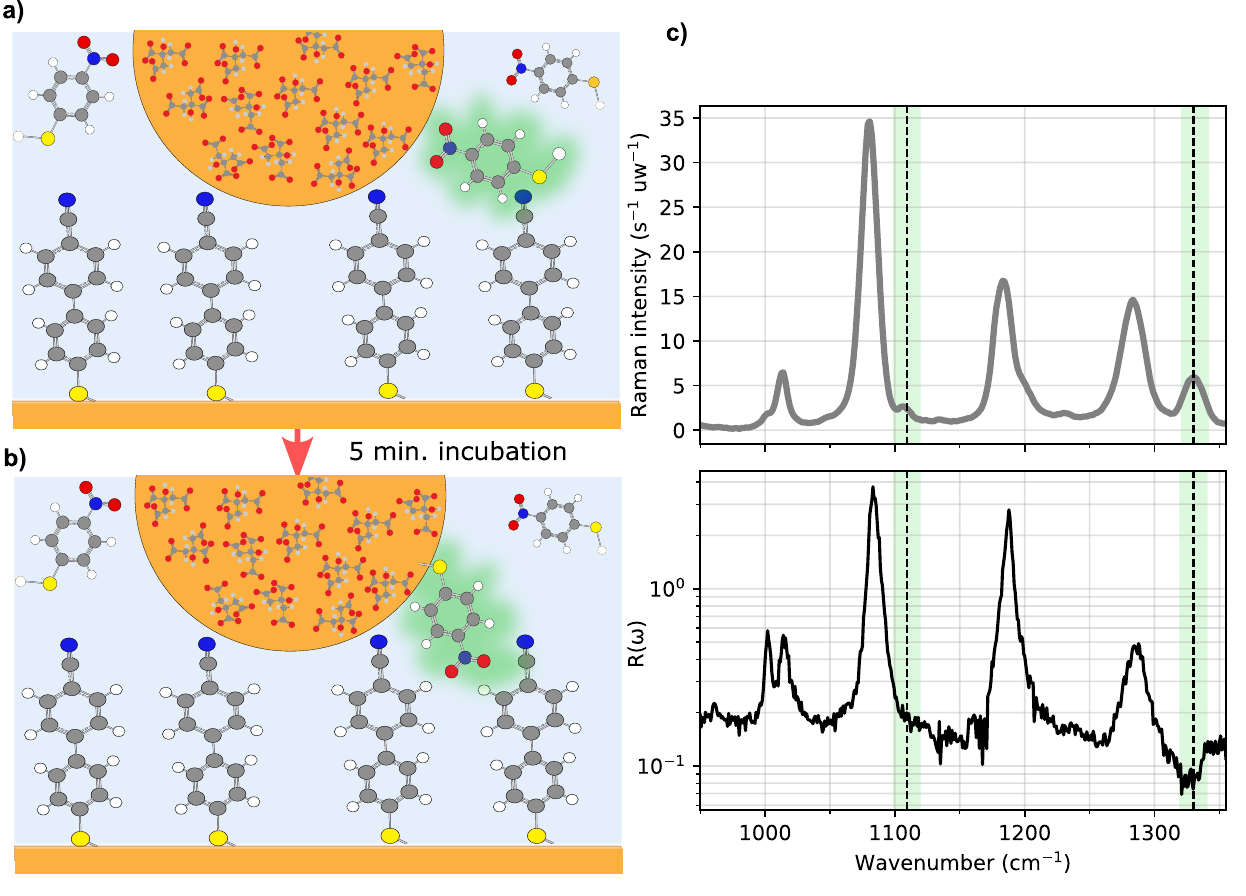}
        \caption{ \textbf{(a)} Sketch of a nanogap formed onto a CN-BPT SAM and immersed in a 1 mM 4NTP ethanol solution for incubation. The side on which 4NTP will bind inside the nanogap is a priori unknown. \textbf{(b)} After 5 minutes of incubation, 4NTP has diffused into the gap. \textbf{(c)} The top panel presents the average SERS spectrum of NPoS cavities after incubation; the dashed lines mark the characteristic 4NTP peaks, which allow us to assess the degree of 4NTP penetration into the gap, but do not reveal any information about its preferential binding surface. The bottom panel displays the average $R(\omega)$ spectrum; from the dip at 1331 cm$^{-1}$, one can infer that the 4NTP molecules are preferentially adsorbed on the nanoparticle.}
        \label{fig:diffusion}
\end{figure}

We conclude this work by demonstrating a potential application of this technique to the field of molecular transport and nanofluidics, where confinement and surface interactions may radically alter molecular properties. 
We perform a molecular diffusion experiment where a pure CN-BPT sample (SAM on the slits) is incubated after the formation of NPoS for 5~min in a 1~mM ethanol solution of 4NTP (Fig.~\ref{fig:diffusion}a,b). The sample is then rinsed with ethanol and dried under a nitrogen flow.
After this treatment, the sample is inspected by SERS and vSFG/vDFG spectroscopy to monitor molecular infiltration into the nanogaps, as reported in Fig.~\ref{fig:diffusion}c. 
Examination of the SERS spectra from a dozen NPoS reveals several levels of 4NTP penetration in the gap, ranging from 5 to 15\%, which reflects the different slit surface roughness and nanoparticle morphology of each NPoS \cite{liuSpatiotemporalRamanProbing2026,wangEffectMirrorQuality2023}. 
For clarity, Fig.~\ref{fig:diffusion}c reports the average spectra of these NPoS. 
The spectrum of $R(\omega)$ presents a marked dip at the characteristic symmetric stretching mode of the nitro group (Fig.~\ref{fig:4NTP}), showing that 4NTP diffuses inside the plasmonic nanogap by prevalently binding to the nanoparticle. 
4NTP thereby displaces the weakly bound citrate molecules, instead of replacing the more strongly bound CN-BPT molecules covering the gold slit. Interestingly, similar infiltration dynamics have been recently confirmed by Goerlitzer \textit{et al.} \cite{goerlitzerSurfaceSelectiveMolecularBinding2026} for phenyl-isocyanide in hybrid Au-Pd NPoMs, using the Raman shift of the isocyanide anchoring group.
The approach presented here, however, does not require a specific anchoring group or tailored metal engineering.

\section*{Conclusion and outlooks}

We demonstrated a fast, robust, non-invasive, and simple method to accurately determine relative molecular orientation inside plasmonic nanogaps, based on ratiometric vDFG/vSFG spectroscopy. 
By probing an individual sub-wavelength nanocavity, the two signals are simultaneously recorded, free of phase-matching constraints. 
The sensitivity of the technique was proven to be better than 60 molecules in the gap, as shown in Fig.~\ref{fig:80nmNPoS}, and should allow single-molecule detection in suitably designed experiments, given the available signal-to-noise ratio. 
Also, we stress that our approach is easily implemented as an add-on to a standard SERS setup, thanks to the commercial availability of powerful and tunable MIR lasers. The SFG and DFG signals are readily observable on simple nanoparticle aggregates with this technique, even without MIR antenna engineering. 
Using this method, we discovered that the background (nonresonant) nonlinear response of the nanogaps is dominated by the metal surface to which the molecules bind and strongly depends on the molecular species, suggesting an important contribution from charge-transfer resonances or related metal-molecule interactions. 
As a potential application to nanofluidics, molecular sensing, and surface chemistry, we monitored molecular transport and preferential absorption in tightly confined plasmonic nanogaps, unambiguously showing that the new molecules bind to the nanoparticle instead of the already functionalized substrate. 


We believe that this method will be useful in better understanding micro/nanofluidics and molecular transport within extremely confined nanochannels. In particular, its straightforward integration with wavelength-multiplexed SERS would enable real-time monitoring of both vertical and lateral infiltration processes in nanofluidic channels \cite{liuSpatiotemporalRamanProbing2026}. 
Recently, several nanofluidic systems such as microplate-on-foil and metal-insulator-metal dual-tone resonators \cite{liuSpatiotemporalRamanProbing2026,palNearFieldVibrationalEnergy} have been demonstrated to possess co-localized MIR and VIS plasmonic resonances, which makes them suitable for the here-presented approach. 
Furthermore, our method could be extended to investigate molecular quantum transport in graphene-sheet or carbon nanotube plasmonic channels, where quantum effects play a significant role \cite{kavokineFluidsNanoscaleContinuum2021}.
Finally, we are convinced that our spectroscopic technique would be beneficial for researcher working with molecular nanojunctions \cite{xuMeasurementSingleMoleculeResistance2003} for electronics and spintronics, a growing field driven by its implication in quantum computing \cite{tanQuantumPlasmonResonances2014,urdampilletaMolecularQuantumSpintronics2011,liuPhotoswitchableQuantumElectrodynamics2023}, nano-catalysis \cite{sanapHotElectronDriven2026} and molecular electronics \cite{limMagneticPlasmonicMultimodular2022,liAtomicallyPreciseConstruction2025}. 
More specifically, monitoring molecular orientation inside molecular junctions could help track geometry rearrangements in ultra-fast photoswitches \cite{hnidMolecularJunctionsTerahertz2024,tangLightDrivenChargeTransport2022,linEngineeringRectificationMolecular2026}, shed light on the reaction pathway within catalytic nanoreactors \cite{zhanPlasmonicNanoreactorsRegulating2021,kianiDistinguishingInnerOuterSphere2024}, or assess the adsorption of different anchoring groups in asymmetric molecular junction. 
In this latter case, the correct positioning of the molecule enables energetic orbital alignment, required for efficient charge conduction through the junction \cite{linEngineeringRectificationMolecular2026,moralesInversionRectifyingEffect2005}.


\section*{Methods}
\subsection{Optical setup}\label{Opticalsetup} 

The continuous-wave monochromatic MIR beam (tunable from 950 to 1650 cm$^{-1}$) is generated by the  MIRcat$^{\rm TM}$ quantum cascade laser and focused onto the sample from the silicon substrate side through a reflective objective (NA 0.78), while a counter-propagating 780 nm laser beam generated by the C-WAVE OPO from Hubner Photonic is focused via a refractive objective (NA 0.9). The generated spontaneous Raman and nonlinear (v-SFG, v-DFG) signals are collected by the same objective employed for 780 nm excitation and directed to a spectrometer (Kymera 193i or
Shamrock 750) through a beam splitter (BS). Notch filters are used to suppress the residual pump. The sample is mounted on a piezo-controlled XYZ stage and measured under ambient conditions.
The wavenumber offsets of Raman spectra were calibrated to match SFG and DFG peaks over the entire MIR laser tuning range. A silicon plate (0.3 mm thick) was introduced in the MIR pathway to attenuate the IR power for the more reactive hybrid sample. IR and VIS powers, employed for results normalization, were measured with a power meter at the focus position, replacing the sample.

\subsection{Slit fabrication}\label{Slit preparation} 
The devices were constructed on 380 $\mu$m thick, double-polished silicon wafers. To ensure good adhesion between Gold and silicon, a 5 nm Aluminum (Al) adhesion layer followed by a 150 nm gold (Au) film was deposited via thermal evaporation, maintained at a precise rate of 0.5 nm/s.
The substrates were then spin-coated with PMMA 950 A4 resist at 2000 rpm and subsequently baked to ensure a uniform resist profile.
Nanoslits were defined using electron-beam lithography (EBL), and after developing the resist, the exposed gold film was etched by a collimated ion beam: a primary etching step was performed at a $10^\circ$ incidence angle to produce V-shaped trenches with slanted sidewalls, followed by a second step at a $10^\circ$° grazing angle to eliminate fencing due to redeposition, ensuring well-defined slit edges. 
The substrates are then left for 24 h in removal 1165 at $60~^\circ$C to strip PMMA; finally, they are immersed in a piranha solution for 10 s to remove organic contaminants and rinsed with water.
The resulting nanoslit geometries featured a consistent length of 1.9 $\mu$m and a width of 140~nm, as confirmed by SEM (Fig.~\ref{fig:SIslits}).

\subsection{Sample preparation}\label{Samplepreparation} 
4-Nitrothiophenol (4NTP) was purchased from BLD Pharma, while 4'-mercaptobiphenylcarbonitrile (CN-BPT) was obtained from Sigma-Aldrich. 
Nanoparticle functionalization was carried out by mixing 900 $\mu$L of citrate-coated Au nanoparticles (1 OD solution from Nanopartz) with 100 $\mu$L of 1 mM thiol solution. The mixture was sonicated for 1 h and then left to rest for an additional hour. It was subsequently centrifuged at 20 g for 10 min to induce nanoparticle precipitation. The supernatant was discarded, and the precipitate was resuspended in deionized water. Redispersion was facilitated by sonication for 10 min.
This washing procedure was repeated two additional times, using ethanol after removal of the supernatant, and finally once more using deionized water.
To verify that no free thiols are present in the solution, after drop-casting the functionalized nanoparticles on the nanoslits, the remaining solution was centrifuged again, the supernatant solution was removed, and the removed solution was dried on a template-stripped Au film. Once dried, citrated-coated Au nanoparticles are drop-cast, and the SERS signal from the so formed NPOM is measured to confirm that no thiols are present.

To prepare self-assembled monolayers (SAM) on the nanoslits, the patterned gold film was first sonicated in acetone for 20 s, then rinsed with ethanol, and finally dried under a nitrogen flow. Immediately after cleaning, the substrate was immersed in a glass vial containing a thiol solution with an overall concentration of 1 mM. The sample was incubated overnight in this molecular solution at ambient temperature.
After incubation, the substrate was rinsed with ethanol, gently sonicated in ethanol for 10 s, rinsed again, and dried under a nitrogen flow. Once dry, an aqueous solution of spherical Au nanoparticles (Nanopartz) was drop-cast onto the substrate. After 1 min, the sample was rinsed with deionized water and dried using a nitrogen gun. 
For the incubation experiments, a CN-BPT SAM NPOS substrate was soaked in a 1mM ethanol solution of 4NTP for 5 min, then the sample was washed with ethanol and dried under nitrogen.

\subsection{Density functional theory (DFT) calculations}

According to the literature, modeling a single thiol molecule by replacing the thiol hydrogen with a gold atom
in the simulation improves the DFT description of the experimental SERS spectra of thiol SAMs, leading to better
agreement between theory and experiment \cite{humbertMultiscaleDescriptionMolecular2012}. All quantum chemistry calculations were performed with Gaussian16.
In our simulations, we considered the gold-linked version (Au–CNBPT), optimizing
their geometries at the def2-TZVPP/B3LYP level with the molecular axis aligned along z. Geometry optimizations
employed tight convergence thresholds (10$^{-5}$ Hartree/Bohr for forces and 4 × 10$^{-5}$ Bohr for RMS displacements,
with maxima $1.5 \times$ larger). Ground-state minima were verified via Hessian analysis.
Raman calculations were carried out employing the aug-cc-pVTZ basis set for H, C, N and S atoms, and aug-cc-pVTZ-PP
for Au, to account for the higher number of polarization and diffusion functions required for such kind of
calculations. We restricted the analysis to the ZZ component of the Raman tensor, since in the nanogap the local field is predominantly normal to the gold surface. 

\section*{Acknowledgments}

This work received funding from the European Union’s Horizon 2020 research and innovation program under Grant Agreement No. 820196 (ERC CoG QTONE) and from the Swiss National Science Foundation (grant number 214993).

\bibliographystyle{ieeetr}

\bibliography{references2}

\end{document}